\documentstyle[preprint,aps]{revtex}

\begin{document}

\draft

%\preprint{xxx-yy/95}

\title{The Absence of Fermionic Superradiance \\
(A Simple Demonstration)}

\author{Hongsu Kim}

\address{Department of Physics\\
Ewha Women's University, Seoul 120-750, KOREA}

\date{May, 1997}

\maketitle

\begin{abstract}
Superradiant scattering, which can be thought of as the
wave analogue of the Penrose process is revisited. As is
well-known, boson fields display superradiance provided
they have frequency in a certain range whereas fermion 
fields do not. A succinct superradiance-checking algorithm
employing particle number or energy current is formally
reviewed and then applied to the case of fermion field.
The demonstrations of the absence of fermionic superradiance
in terms of the particle number current exist in the
literature but they are in the context of two-component
SL(2,C) spinor formalism for massive spinor and SO(3,1)
Dirac spinor formalism for massless spinor. 
Here we present an alternative
demonstration in terms of both particle number and energy
current but in a different context of local SO(3,1) Dirac 
spinor formalism for both massless and massive spinors.
It appears that our presentation confirms the
absence of fermionic superradiance in a more simple and
systematic manner.
\end{abstract}


\narrowtext
%\twocolumn

%%%%%%%%%%%%%%%%%%%%%%%%%%%%%%%%%%%%%%%%%%%%%%%%%%%%%%%%%%%%%%%%%%%%%%%%%%%%%%%

%\newpage

%%%%%%%%%%%%%%%%%%%%%%%%%%%%%%%%%%%%%%%%%%%%%%%%%%%%%%%%%%%%%%%%%%%%%%%%%%%%%%%%%%%%%%

\centerline {\bf I. Introduction}

A black hole is, by definition, a ``region of no escape". 
No massive object
or even the massless light ray, therefore,  can ever be extracted from a 
black hole. When it comes to rotating black holes such as the Kerr
family of solutions, however, things are not so simple and indeed energy 
can be extracted from black holes as was first noted by Penrose [1].  
Briefly, this energy extraction mechanism proposed by Penrose and 
hence is called ``Penrose process" [1,5] can be understood as follows.
In Kerr geometry, the surface on which $g_{tt}$ vanishes does not
coincide with the event horizon exept at the poles. The toroidal space
inbetween the two surfaces is called ``ergosphere" and in particular the
outer boundary of this ergosphere on which $g_{tt}$ vanishes is dubbed
``static limit" because it can be seen that inside of which no observer 
can possibly remain static. Namely the time translational Killing field 
$\xi^{\mu} = (\partial/\partial t)^{\mu}$ 
becomes spacelike inside the ergosphere
and so does the conserved component $p_{t}$ of the four momentum.
As a consequence, the energy of a particle in this ergoregion, as 
perceived by an observer at infinity, can  be negative. This last
fact leads to a peculiar possibility that, in principle, one can
devise a physical process which extracts energy and angular momentum from
the black hole. The Penrose process, however, requires a precisely
timed breakup of the incident particle at the relativistic velocities
and thus is not a very practical energy extraction scheme.
Perhaps because of this reason, an alternative study of energy
extraction mechanism, known as ``superradiant scattering" [2,3,4,5]
was considered.
In a sense, it can be thought of as a wave analogue of the Penrose process.
If a wave is incident upon a black hole, the part of the wave (``transmitted"
wave) will be absorbed by the black hole and the part of the wave
(``reflected" wave) will escape back to infinity. Normally, the transmitted
wave will carry positive energy into the black hole and the reflected wave
will have less energy than the incident wave. However, for a scalar wave
with the time ($t$) and azimuthal angle ($\phi$) dependence given by
$e^{i(m\phi - \omega t)}$ (with $m$ and $\omega$ being
the azimuthal number and the frequency respectively), the transmitted
wave will carry negative energy into the black hole and the reflected wave
will escape to infinity with greater amplitude and energy than the
originally incident one provided the scalar wave has the frequency in
the range [2,4,5]
\begin{eqnarray}
0 < \omega < m\Omega_{H}  \nonumber
\end{eqnarray}
where $\Omega_{H}$ denotes the angular velocity of the rotating hole
at the event horizon. The ``scalar waves" such as electromagnetic and
gravitational waves exhibit this superradiance [2,4] when they have frequency
in the range given above. Curiously enough, it is known that fermion
fields do not display superradiance [3,5]. Conventionally, the typical way
of demonstrating the presence or absence of  superradiance is to
define the reflection and transmission coefficients in terms of the
solutions of the wave (field) equations in the background of the
Kerr spacetime and then see if the reflection coefficient
can exceed unity [4]. And unlike the case of scalar waves, the reflection
coefficient in the case of fermion field never exceeds unity for all
values of the frequency including the ones in the range given above [4].
In the present work, we would like to present a simple yet very
solid formalism which demonstrates the absence of the superradiance
in the case of massive or massless fermion field. It involves 
standard formulation of spinor field theory in curved background
spacetime (the Kerr black hole background for
our case) [6] associated with the Riemann-Cartan formulation of general 
relativity (which 
utilizes the soldering form particularly suitable for the case of spinor
field). Thus in this formulation, we need to put the Kerr metric
given in Boyer-Lindquist coordinates [7] in ADM's (3+1) space-plus-time
split form to extract the soldering form (i.e., the non-coordinate
basis 1-form).
This, then, allows us to show that
the energy current or the particle number current 
(which will be defined shortly in the next section) [5] of spinor field
flowing into the black hole through the event horizon exactly
vanishes establishing the absence of superradiance. 
The demonstration employing the fermionic particle number current may not 
be new [3]. But the existing works in the literature are all in the
context of two-component SL(2,C) spinor formalism for massive spinor
and SO(3,1) Dirac spinor formalism for massless spinor.
Our formalism that we shall present here works with Dirac spinor
(but in a different context from the existing work)
associated with the local SO(3,1) group for both massless and massive
spinors and it allows us to compute both
the energy flux and the particle number flux in a more systematic
and straightforward manner.
\\
\centerline {\bf II. General Formalism}
\\
In this section, we would like to first set up a general formalism [5]
that allows us to determine whether or not the superradiance is
actually present in the case of scalar or fermion field. Then later
on we shall illustrate, as an exmple,  the presence of superradiance 
in the case of scalar (complex scalar) field. \\
To this end, we introduce two quantities of central importance,
``energy current" and ``particle number current".
First, we begin with the energy current. Generally, the ``energy
current" of a field in curved background spacetime is defined by [5]
\begin{eqnarray}
J_{\mu} \equiv - T_{\mu\nu} \xi^{\nu}
\end{eqnarray}
with $\xi^{\mu} = (\partial/\partial t)^{\mu}$ being the time
translational Killing field of a stationary, axisymmetric 
spacetime (which is the Kerr black hole spacetime
for our case). This quantity is obviously conserved owing to
the energy-momentum conservation and the Killing equation
$\nabla^{\mu}\xi^{\nu} + \nabla^{\nu}\xi^{\mu} = 0$
satisfied by the Killing field $\xi^{\mu}$, i.e.,
\begin{eqnarray}
\nabla^{\mu}J_{\mu} = - (\nabla^{\mu}T_{\mu\nu})\xi^{\nu}
- T_{\mu\nu}(\nabla^{\mu}\xi^{\nu}) = 0. \nonumber
\end{eqnarray}
Next, we turn to the particle number current. Generally
speaking, for field theories with action possessing the
global U(1) transformation (i.e., phase transformation)
symmetry, (e.g. complex scalar field theory and fermion
field theory) the associated Noether current can be identified
with the particle number current. Namely the Noether current
of the typical form
\begin{eqnarray}
j^{\mu} = {\delta {\cal L}\over \delta (\nabla_{\mu}
\phi^{i})} \delta \phi^{i} 
\end{eqnarray}
(where ${\cal L}$ denotes the Lagrangian density) is defined
to be the particle number density. Then this particle number
density is covariantly conserved as well due to the
Euler-Lagrange's equation of motion and the invariance of the
action, $\nabla_{\mu}j^{\mu} = 0$. \\
Now in order eventually to determine the presence or absence of
the superradiant scattering, we consider a region $K$ of
spacetime of which the boundary consists of two spacelike
hypersurfaces $\Sigma_{1}$ at ($t$) and  $\Sigma_{2}$ at 
($t+\delta t$) (the constant time slice  $\Sigma_{2}$
is a time translate of  $\Sigma_{1}$ by $\delta t$) and
two timelike hypersurfaces $H$ (black hole horizon at
$r = r_{+}$) and $S_{\infty}$ (large sphere at spatial infinity 
$r \rightarrow \infty$). The appropriate directions of the 
hypersurface normal vector $n^{\mu}$ on each part of the 
boundary are ; $n^{\mu}$ is future-directed on $\Sigma_{1}$,
past-directed on  $\Sigma_{2}$. It is pointing inward the
black hole on the event horizon $H$ and pointing outward to
infinity on  $S_{\infty}$. Then next, consider integrating
the quantity $\nabla^{\mu}J_{\mu}$ (which leads to the
``energy flux" crossing each part of the boundary upon
utilizing the Gauss's theorem) or  $\nabla_{\mu}j^{\mu}$
(which leads to the ``particle number current") over the
region $K$ of spacetime. By using Gauss's theorem we have
\begin{eqnarray}
0 &=& \int_{K} d^4x \sqrt{g} \nabla_{\mu}j^{\mu} \nonumber \\
&=& \int_{\partial K} d^3x \sqrt{h} n_{\mu}j^{\mu} \\
&=& \int_{\Sigma_{1}(t)}n_{\mu}j^{\mu} + \int_{\Sigma_{2}(t+\delta t)} 
n_{\mu}j^{\mu} + \int_{H(r_{+})} n_{\mu}j^{\mu} + 
\int_{S_{\infty}} n_{\mu}j^{\mu} \nonumber
\end{eqnarray}
where $h_{\mu\nu}$ denotes the 3-metric induced on the boundary
$\partial K$ of the region $K$.
Now the terms in the last line of eq.(3) need some explanations.
For boson or fermion field with time dependence of the form
$e^{-i\omega t}$ (which we shall assume throughout) the first two
terms cancel with each other by time translation symmetry.
The third term represents the net particle number flow into the rotating 
black
hole while the last term stands for the net particle number flow out of $K$
to infinity, i.e., the outgoing minus incoming particle number
through $S_{\infty}$ during the time $\delta t$. Thus we end up with
the result 
\begin{eqnarray}
\int_{S_{\infty}} n_{\mu}j^{\mu} = - \int_{H(r_{+})} n_{\mu}j^{\mu}
\end{eqnarray}
which states that the {\it net particle number flow out of $K$ or the
``outgoing minus incoming particle number" equals ``minus" of the
net particle number flow into the rotating black hole}.
Therefore, now we can establish the criterion for the occurrence of
superradiant scattering ; If the quantity on the right hand side
$\int_{H(r_{+})} n_{\mu}j^{\mu}$, namely the net particle number flowing
down the hole, is negative (zero or positive), it means that the
outgoing particle number flux is greater (smaller) than the incident
one and hence the superradiance is present (absent). \\
Thus far we have established the criterion for the occurrence of
superradiance in terms of the ``particle number current" $j^{\mu}$.
An equivalent criterion can be derived in terms of the ``energy 
current" $J_{\mu}$ if we replace  $n_{\mu}j^{\mu}$ by
$<n^{\mu}J_{\mu}>$ (where $<...>$ denotes time averaged quantity)
and replace ``particle number current" with ``energy current"
respectively in the above formalism. \\
Also note that the hypersurface normal $n^{\mu}$ on the black hole
event horizon $H$ is pointing inward the hole and hence is opposite
to the direction of the Killing field [5]
\begin{eqnarray}
\chi^{\mu} = \xi^{\mu} + \Omega_{H} \psi^{\mu}
\end{eqnarray}
which is outer normal to the rotating hole's event horizon
(here $\xi^{\mu}=(\partial/\partial t)^{\mu}$ and 
$\psi^{\mu}=(\partial/\partial \phi)^{\mu}$ denotes Killing
fields associated with the time translational and rotational 
isometries of the stationary, axisymmetric Kerr black hole
spacetime respectively and $\Omega_{H}$ denotes the angular velocity
of the event horizon of the hole). \\
Thus our task of checking the presence or absence of superradiance
reduces to the computation of the net particle number (or energy)
current flowing into the rotating hole through its event horizon
\begin{eqnarray}
\int_{H(r_{+})} n_{\mu}j^{\mu} = - \int_{H(r_{+})} \chi_{\mu}j^{\mu}.
\end{eqnarray}
Now, before we demonstrate the absence of the superradiance in the case
of fermion field in the next section, it will be instructive to
illustrate the presence of the superradiance in a boson field case
using the superradiance-checking formalism introduced above. \\
Thus consider a complex scalar field theory in a stationary, axisymmetric
background spacetime (which we take to be the Kerr black hole 
geometry) described by the action [6] (here in this work, we employ the
Misner-Thorne-Wheeler sign convention [8] in which the metric has the
sign of $g_{\mu\nu} = diag (- + + +)$)
\begin{eqnarray}
S = - \int d^4x \sqrt{g} [\nabla_{\mu}\Phi^{\ast} \nabla^{\mu}\Phi +
(M^2 + \xi R)\Phi^{\ast}\Phi]
\end{eqnarray}
and the classical field equations
\begin{eqnarray}
&\nabla_{\mu}&\nabla^{\mu}\Phi - (M^2 + \xi R)\Phi = 0, \\
&\nabla_{\mu}&\nabla^{\mu}\Phi^{\ast} - (M^2 + \xi R)\Phi^{\ast} = 0 
\nonumber
\end{eqnarray}
where $\xi$, $M$ and $R$ denotes some constant (for example, $\xi =
1/6$ with $M = 0$ corresponds to ``conformal couplig''), 
the mass of the scalar 
field and the scalar curvature of the background spacetime respectively.
Then we consider a situation when a complex scalar wave with particular
frequency
\begin{eqnarray}
\Phi(x) = \Phi_{0}(r,\theta)e^{i(m\phi -\omega t)} 
\end{eqnarray}
is incident on and reflected by the Kerr black hole.
Since the Lagrangian density of this complex scalar field 
in eq.(7) is invariant
under the global U(1) (or phase) transformation
\begin{eqnarray}
\Phi(x) &\rightarrow & e^{-i\alpha}\Phi(x), \nonumber \\
\Phi^{\ast}(x) &\rightarrow & e^{i\alpha}\Phi^{\ast}(x) \nonumber
\end{eqnarray}
corresponding Noether current exists and it is
\begin{eqnarray}
j^{\mu} = -i (\Phi^{\ast} \nabla^{\mu}\Phi - \Phi \nabla^{\mu}\Phi^{\ast}).
\end{eqnarray}
This Noether current is the particle number current and it can be seen to
be conserved owing to the classical field equations in eq.(8)
\begin{eqnarray}
\nabla_{\mu}j^{\mu} = 0. \nonumber
\end{eqnarray}
According to the criterion for the occurrence of the superradiance stated
earlier, all we have to do now is to evaluate the net particle number
flowing into the black hole, $\int_{H(r_{+})}n_{\mu}j^{\mu}$ and see if it 
can be negative. Thus on the horizon $r = r_{+}$, we compute the particle
number flux and it is
\begin{eqnarray}
n_{\mu}j^{\mu} &=& - \chi^{\mu}j_{\mu} \nonumber \\
&=& i (\Phi^{\ast}\chi^{\mu}\nabla_{\mu}\Phi - 
\Phi \chi^{\mu}\nabla_{\mu} \Phi^{\ast}) \\
&=& i [\Phi^{\ast}({\partial \over \partial t} + \Omega_{H}
{\partial \over \partial \phi})\Phi -
\Phi ({\partial \over \partial t} + \Omega_{H}
{\partial \over \partial \phi})\Phi^{\ast}] \nonumber \\
&=& 2 (\omega - m\Omega_{H})\mid \Phi_{0} \mid^{2}. \nonumber
\end{eqnarray}
Thus for a complex scalar field with frequency in the range
\begin{eqnarray}
0 < \omega < m\Omega_{H}
\end{eqnarray}
the net particle number flowing down the hole is negative and hence
\begin{eqnarray}
\int_{S_{\infty}} n_{\mu}j^{\mu} =
- \int_{H(r_{+})}  n_{\mu}j^{\mu} > 0
\end{eqnarray}
namely the outgoing minus incident particle number flux through the
large sphere $S_{\infty}$ is positive indicating the occurrence
of superradiance in the case of a scalar field.
Before we end, we would like to comment on the following point.
In the process of drawing the conclusion that the scalar field
we considered displays superradiance provided it has the frequency
in the range given above, almost no reference has been made to the 
specifics of the background spacetime geometry, i.e., the Kerr 
geometry. The only ingredient associated with the Kerr
geometry entered the analysis was the use of the Killing field
$\chi^{\mu}$ which is the outer normal to the rotating hole's
event horizon. Thus we really need not know even the concrete form of
the metric of Kerr spacetime. In a sense, this point may
imply that the occurrence of superradiance in boson field case
is an unshakeable, solid phenomenon but in another sense it
makes us feel rather uncomfortable. In the case of fermion field we
shall discuss in the following section, situation changes ;
one needs the specifics of the Kerr geometry to reach 
the conclusion on the absence of superradiance, and we feel that
this seems more natural.
\\
\centerline {\bf III. Absence of superradiance in the case of fermion field}
\\
As already mentioned and as we shall see shortly as well, in
order to demonstrate the absence of superradiance in the 
fermion field case, one needs the concrete geometry structure
of the background Kerr black hole. Besides, the standard
formulation of spinor field theory in curved background spacetime
is associated with the Riemann-Cartan formulation 
of general relativity in which one of the basic
computational tools is the use of the non-holonomic basis 1-form
(i.e., ``soldering form"). Thus here we begin by casting the
Kerr metric (given in Boyer-Lindquist coordinates [7]) into
the ADM's (3+1) space-plus-time split form which proves to be
suitable to be converted to the one in non-coordinate basis
\begin{eqnarray}
ds^2 &=& -N^2 dt^2 + h_{rr}dr^2 + h_{\theta \theta}d\theta^2 +
h_{\phi \phi}[d\phi + N^{\phi}dt]^2 \nonumber \\
&=& g_{\mu\nu}dx^{\mu}dx^{\nu} = \eta_{ab}e^{a}e^{b} 
\end{eqnarray}
where Greek indices refer to the accelerated frame of reference 
(i.e., coordinate basis, $\mu = t,~r,~\theta,~\phi$) and
the Roman indices refer to the locally inertial reference frame
(i.e., non-coordinate basis, $a = 0,~1,~2,~3$). 
Also we used the definitions for
the soldering form (``vierbein") $g_{\mu\nu} = \eta_{ab}
e^{a}_{\mu}e^{b}_{\nu}$ and the non-coordinate basis 1-form
$e^{a} = e^{a}_{\mu}dx^{\mu}$. In the ADM's (3+1) split form
above, the lapse, shift functions and the spatial metric components
are given respectively by
\begin{eqnarray}
N^2(r,\theta) &=& [{\Delta - a^2\sin^2\theta \over \Sigma}] +
R^{-2}(r,\theta)[{r^2+a^2 - \Delta \over \Sigma}]^2 a^2\sin^4\theta,
\nonumber \\
N^{\phi}(r,\theta)&=& - R^{-2}(r,\theta)[{r^2+a^2 - \Delta \over \Sigma}]
a\sin^2\theta, ~~~N^{r} = N^{\theta} = 0, \nonumber \\
h_{rr}(r,\theta) &=& f^{-2}(r,\theta) = {\Sigma \over \Delta},
~~~h_{\theta \theta}(r,\theta) = g^{2}(r,\theta) = \Sigma , \\
h_{\phi \phi}(r,\theta) &=& R^2(r,\theta) = 
[{(r^2+a^2)^2 - \Delta a^2\sin^2\theta \over \Sigma}]\sin^2\theta,
\nonumber \\
h_{r\theta} &=& h_{\theta r} = 0, ~~~h_{r\phi} = h_{\phi r} = 0,
~~~h_{\theta \phi} = h_{\phi \theta} = 0 \nonumber
\end{eqnarray}
where $\Sigma = r^2 + a^2 \cos^2\theta$ and 
$\Delta = r^2 - 2M_{KN}r + a^2$ with $M_{KN}$  denoting
the mass of the Kerr black hole respectively. As is well known the
event horizon developes at the larger zero of $\Delta(r_{+}) = 0$.\\
Actually, the virtue of writing the Kerr metric in the ADM's
(3+1) split form as in eq.(14) above is that from which now one can 
read off the 
non-coordinate basis 1-form easily as follows
\begin{eqnarray}
e^{0} &=& e^{0}_{\mu}dx^{\mu} = N dt, \nonumber \\
e^{1} &=& e^{1}_{\mu}dx^{\mu} = \sqrt{h_{rr}} dr = f^{-1} dr, \\
e^{2} &=& e^{2}_{\mu}dx^{\mu} = \sqrt{h_{\theta \theta}} d\theta =
g d\theta \nonumber \\
e^{3} &=& e^{3}_{\mu}dx^{\mu} = \sqrt{h_{\phi \phi}}(d\phi + N^{\phi}dt)
= R(d\phi + N^{\phi}dt). \nonumber
\end{eqnarray}
Equvalently, the vierbein and the inverse vierbein can be read off as
\begin{equation}
e^{a}_{\mu} = \left( \begin{array}{cccc}
                              N & 0 & 0 & 0 \\
                              0 & f^{-1} & 0 & 0 \\
                              0 & 0 & g & 0 \\
                              RN^{\phi} & 0 & 0 & R \\ 
                     \end{array} \right),
~~~e^{\mu}_{a} = \left( \begin{array}{cccc}
                              N^{-1} & 0 & 0 & 0 \\
                              0 & f & 0 & 0 \\
                              0 & 0 & g^{-1} & 0 \\
                              -N^{-1}N^{\phi} & 0 & 0 & R^{-1} \\
                     \end{array} \right)
\end{equation}
Further, the spin connection 1-form $\omega^{ab} = \omega^{ab}_{\mu}
dx^{\mu}$ can be obtained from the Cartan's 1st structure equation,
$de^{a} + \omega^{a}_{b}\wedge e^{b} = 0$ using the non-coodinate
basis 1-form given in eq.(16).  Here, however, we do not
look for the spin connection since we shall not really need its
explicit form in the discussion below leading to the conclusion on
the absence of fermionic superradiance. \\
Now, for later use we write the $\gamma$-matrices with coordinate basis
indices in accelerated frame (i.e., in Boyer-Lindquist coordinates)
in terms of those with non-coordinate basis indices in locally-inertial
frame using the soldering form (inverse vierbein) given above, i.e.,
$\gamma^{\mu}(x) = e^{\mu}_{a}(x)\gamma^{a}$
\begin{eqnarray}
\gamma^{t} &=&  e^{t}_{a}\gamma^{a} = N^{-1}\gamma^{0}, \nonumber \\
\gamma^{r} &=&  e^{r}_{a}\gamma^{a} = f\gamma^{1}, \\
\gamma^{\theta} &=&  e^{\theta}_{a}\gamma^{a} = g^{-1}\gamma^{2},
\nonumber \\
\gamma^{\phi} &=&  e^{\phi}_{a}\gamma^{a} = -N^{-1}N^{\phi}\gamma^{0}
+ R^{-1}\gamma^{3}. \nonumber
\end{eqnarray}
With this preparation, now we consider the spinor field theory in
the background of this Kerr spacetime described by the action [6]
\begin{eqnarray}
S = \int d^4x \sqrt{g} \{ {i\over 2}[\bar{\psi}\gamma^{\mu}
\overrightarrow{\nabla}_{\mu}\psi - \bar{\psi}\gamma^{\mu}
\overleftarrow{\nabla}_{\mu}\psi] - M \bar{\psi}\psi \}
\end{eqnarray}
and the classical field equations, i.e., curved spacetime Dirac equations
\begin{eqnarray}
(i\gamma^{\mu}\overrightarrow{\nabla}_{\mu} - M)\psi &=& 0, \\
\bar{\psi}(i\gamma^{\nu}\overleftarrow{\nabla}_{\nu} + M) &=& 0 
\nonumber 
\end{eqnarray}
where $\gamma^{\mu}(x) = e^{\mu}_{a}(x)\gamma^{a}$ is the curved
spacetime $\gamma$-matrices obtained in eq.(18) above satisfying
$\{ \gamma^{\mu}(x), ~\gamma^{\nu}(x)\} = - 2g^{\mu \nu}(x)$ with
$e^{a}_{\mu}$ and $e^{\mu}_{a}$ being the vierbein and its inverse
as obtained in eq.(17) and they are defined by 
$g_{\mu\nu}(x) = \eta_{ab}e^{a}_{\mu}(x)e^{b}_{\nu}(x)$ and
$e^{a}_{\mu}e^{\mu}_{b} = \delta^{a}_{b}$,
$e^{\mu}_{a}e^{a}_{\nu} = \delta^{\mu}_{\nu}$.
And $\nabla_{\mu} = [\partial_{\mu} - {i\over 4}\omega^{ab}_{\mu}(x)
\sigma_{ab}]$ is the covariant derivative with $\omega^{ab}_{\mu}(x)$
being the spin connection (that can, as mentioned, be obtained from the 
vierbein in eq.(17)) and $\sigma_{ab} = {i\over 2} [\gamma_{a},~\gamma_{b}]$ 
being the SO(3,1) group generator in the spinor representation. \\
Again, we consider a situation when a spinor wave with particular
frequency
\begin{eqnarray}
\psi(x) = u(p,s)\psi_{0}(r,\theta)e^{i(m\phi -\omega t)}
\end{eqnarray}
is incident on and reflected by the Kerr black hole.
Here $s$ denotes spin state, $p^{t} = \omega$ and the 4-component
Dirac spinor $u(p,s)$ satisfies the Dirac equation in curved
spacetime given above.
Since the Lagrangian density of this spinor field 
given in eq.(19) is also 
invariant under the global U(1) (or phase) transformation
\begin{eqnarray}
\psi(x) &\rightarrow & e^{-i\alpha}\psi(x), \nonumber \\
\bar{\psi}(x) &\rightarrow & \bar{\psi}(x)e^{i\alpha} \nonumber
\end{eqnarray}
corresponding Noether current exists and it is
\begin{eqnarray}
j^{\mu} &=& {\delta {\cal L}\over 
\delta (\overrightarrow{\nabla}_{\mu}\psi)}\delta \psi
+ \delta \bar{\psi} 
{\delta {\cal L}\over 
\delta (\bar{\psi}\overleftarrow{\nabla}_{\mu})} \nonumber \\
&=& \bar{\psi} \gamma^{\mu} \psi. 
\end{eqnarray}
This Noether current is identified with the particle number current
and it can be seen to be conserved due to the Dirac equations given above
\begin{eqnarray}
\nabla_{\mu}j^{\mu} &=& \bar{\psi}\gamma^{\mu}\overleftarrow{\nabla}_{\mu}
\psi + \bar{\psi}\gamma^{\mu}\overrightarrow{\nabla}_{\mu}\psi \nonumber \\
&=& iM\bar{\psi}\psi -  iM\bar{\psi}\psi = 0.
\end{eqnarray}
An alternative way of deriving the number current and confirming its
covariant conservation is to multiply two Dirac equations in eq.(20)
by $\bar{\psi}$ and $\psi$ respectively and add them up to yield eq.(23).
According to the criterion for checking the occurrence of the superradiance 
stated earlier in the previous section, all we have to do is to evaluate 
the net particle number
flowing into the black hole, $\int_{H(r_{+})}n_{\mu}j^{\mu}$ and see if it
can be negative. Thus on the horizon $r = r_{+}$, we compute the 
fermionic particle number flux and it is
\begin{eqnarray}
n_{\mu}j^{\mu} &=& - \chi^{\mu}j_{\mu} = - \bar{\psi} g_{\alpha\beta}
\chi^{\alpha}\gamma^{\beta}\psi \nonumber \\
&=&  - \bar{\psi} g_{\alpha\beta} (\delta^{\alpha}_{t} + 
\Omega_{H}\delta^{\alpha}_{\phi}) \gamma^{\beta} \psi \\
&=& - \bar{\psi}[(g_{tt} + \Omega_{H}g_{t\phi})\gamma^{t} +
(g_{t\phi} + \Omega_{H}g_{\phi\phi})\gamma^{\phi}]\psi \nonumber \\
&=& - \bar{\psi}[ - N\gamma^{0} + R (N^{\phi} + \Omega_{H})\gamma^{3}]
\psi = 0. \nonumber
\end{eqnarray}
where we used the relation between $\gamma$-matrices with coordinate basis
indices in accelerated frame (i.e., in Boyer-Lindquist coordinates)
and those with non-coordinate basis indices in locally-inertial
frame, $\gamma^{\mu}(x) = e^{\mu}_{a}(x)\gamma^{a}$ derived earlier
in eq.(18) and $g_{tt} = -[N^2-R^2(N^{\phi})^2]$, $g_{t\phi} = R^2N^{\phi}$
and $g_{\phi\phi} = R^2$. And to get the last equality to zero we used
\begin{eqnarray}
N^2(r_{+},~\theta) = 0, ~~~N^{\phi}(r_{+},~\theta) = -({a\over r^2_{+} +
a^2}) = - \Omega_{H}. \nonumber
\end{eqnarray}
This result indicates that the net fermionic particle number flux flowing 
down the hole through its event horizon is zero {\it irrespective of
the frequency of the fermion field and regardless of its being massive
or massless}. Therefore the outgoing minus incident fermionic particle
number flux through the large sphere $S_{\infty}$ is zero
\begin{eqnarray}
\int_{S_{\infty}} n_{\mu}j^{\mu} = - \int_{H(r_{+})}n_{\mu}j^{\mu} = 0
\nonumber
\end{eqnarray}
establishing the absence of superradiance in the case of fermion field. \\
Thus far we have illustrated the absence of fermionic superradiance in 
terms of the particle number flux. One can draw the same conclusion in
terms of the energy flux we introduced earlier by showing that it also
is zero through the event horizon of the Kerr black hole.
Thus in the following we shall do this. \\
As mentioned earlier, the energy current is defined to be
\begin{eqnarray}
J_{\mu} = - T_{\mu\nu}\xi^{\nu} \nonumber
\end{eqnarray}
with $\xi^{\mu}$ again being the time translational Killing field
and this energy current is indeed conserved due to the energy-momentum
conservation and the Killing equation satisfied by  $\xi^{\mu}$.
For a fermion field described by the action given earlier, the 
energy-momentum tensor is given by [6]
\begin{eqnarray}
T^{\mu\nu} &=& {1\over 2(det ~e)}\eta^{ab} [e^{\mu}_{a}{\delta S\over
\delta e^{b}_{\nu}} + e^{\nu}_{b}{\delta S\over\delta e^{a}_{\mu}}] \\
&=& {i\over 4}\{ [\bar{\psi}\gamma^{\mu}
\overrightarrow{\nabla}^{\nu}\psi - \bar{\psi}\gamma^{\mu}
\overleftarrow{\nabla}^{\nu}\psi] 
+  [\bar{\psi}\gamma^{\nu}
\overrightarrow{\nabla}^{\mu}\psi - \bar{\psi}\gamma^{\nu}
\overleftarrow{\nabla}^{\mu}\psi] \}. \nonumber
\end{eqnarray}
Now in exactly the same manner as we worked with the particle number
current, we evaluate the net ``time averaged" energy current
flowing into the black hole, $\int_{H(r_{+})}< n^{\mu}J_{\mu}>$ and see 
if it can be negative, but in this time using the form of the spinor wave
with particular frequency given by 
\begin{eqnarray}
\psi(x) &=& u(p,s)Re[\psi_{0}(r,\theta) e^{i(m\phi - \omega t)}], \\
\bar{\psi}(x) &=& \bar{u}(p,s)Re[\psi^{\ast}_{0}(r,\theta)
e^{-i(m\phi - \omega t)}]. \nonumber
\end{eqnarray} 
Thus on the horizon $r = r_{+}$, we compute the
fermionic time averaged energy flux and it is
\begin{eqnarray}
<n^{\mu}J_{\mu}> &=& - <\chi^{\mu}J_{\mu}> = 
<T_{\mu\nu}\chi^{\mu}\xi^{\nu}> \\
&=& < {i\over 4}\{ [\bar{\psi}\chi^{\mu}\gamma_{\mu}
\xi^{\nu}\overrightarrow{\nabla}_{\nu}\psi - 
\bar{\psi}\chi^{\mu}\gamma_{\mu}
\xi^{\nu}\overleftarrow{\nabla}_{\nu}\psi] \nonumber \\
&+&  [\bar{\psi}\xi^{\nu}\gamma_{\nu}
\chi^{\mu}\overrightarrow{\nabla}_{\mu}\psi - 
\bar{\psi}\xi^{\nu}\gamma_{\nu}
\chi^{\mu}\overleftarrow{\nabla}_{\mu}\psi] \} > = 0 \nonumber
\end{eqnarray}
where we used the facts that $\chi^{\mu}\gamma_{\mu} = 0$
on the event horizon as we have shown in eq.(24) above 
in the evaluation
of $n_{\mu}j^{\mu}$ and that the two terms on the left hand side
of the last line exactly cancel with each other since
$\xi^{\nu}\gamma_{\nu} = (g_{tt}\gamma^{t} + g_{t\phi}\gamma^{\phi})
= [-N(r_{+})\gamma^{0} + R(r_{+})N^{\phi}(r_{+})\gamma^{3}] =
R(r_{+})N^{\phi}(r_{+})\gamma^{3}$ and
\begin{eqnarray}
\chi^{\mu}\overrightarrow{\nabla}_{\mu}\psi &=& 
[(\omega - m\Omega_{H})Y_{0}(r_{+},\theta)\cos (m\phi - \omega t)
- (\omega + m\Omega_{H})X_{0}(r_{+},\theta)\sin (m\phi - \omega t)]u(p,s)
\nonumber \\
&-& {i\over4}(\omega^{ab}_{t}+\Omega_{H}\omega^{ab}_{\phi})\sigma_{ab}\psi,
\nonumber \\
\bar{\psi}\chi^{\mu}\overleftarrow{\nabla}_{\mu} &=& 
[(\omega - m\Omega_{H})Y_{0}(r_{+},\theta)\cos (m\phi - \omega t)
- (\omega + m\Omega_{H})X_{0}(r_{+},\theta)\sin (m\phi - \omega t)]
\bar{u}(p,s) \nonumber \\
&-& {i\over4}\bar{\psi}(\omega^{ab}_{t}+\Omega_{H}\omega^{ab}_{\phi})
\sigma_{ab} \nonumber
\end{eqnarray}
where $X_{0}(r,\theta) = Re[\psi_{0}(r,\theta)]$ and
$Y_{0}(r,\theta) = Im[\psi_{0}(r,\theta)]$.
Again, the net time averaged fermionic energy flux flowing
down the hole through its event horizon is zero irrespective of
the frequency of the fermion field and regardless of its being massive
or massless. Therefore the outgoing minus incident time averaged fermionic 
energy flux through the large sphere $S_{\infty}$ is zero confirming the
absence of fermionic superradiance.
\\
\centerline {\bf IV. Discussions}
\\
Now we end with some comments.
It is well-known that the behavior of both boson and fermion fields
incident upon a rotating Kerr black hole is in close analogy
to a famous effect in relativistic quantum mechanics known as the ``Klein 
paradox". Recall that if a Klein-Gordon field in one spatial dimension
is incident upon an electrostatic potential $V$, the reflected wave
emerge with greater amplitude and energy than the incident one 
provided certain condition is met. For a Dirac field, however,
the reflected wave will be smaller than the incident one.
In quantum field theory, the
interpretation of the Klein paradox is that in both the boson and
fermion cases, particle-antiparticle pairs are spontaneously created
in the strong electrostatic field associated with the potential V.
And when incoming particles also are present, stimulated emision
occurs in the boson case and in the classical limit, this results in
the amplified reflected wave obtained with the classical analysis.
The close analogy between the Klein paradox and the superradiant
scattering of waves by a Kerr black hole suggests that
spontaneous particle creation should occur near the Kerr
black hole horizon. Indeed this is the case and it is directly
related to the origin of Hawking evaporation of black holes [9,6].\\
Thus far the standard, traditional way of demonstrating the
presence of superradiance in boson field case and the absence
in fermion field case has been to define the reflection and
transmission coefficients in terms of the solutions of the wave 
equations in the Kerr black hole background and see if
the reflection coefficient can exceed unity under certain
circumstances [4]. The demonstration in terms of 
the fermionic particle number current in the context of 
two-component SL(2,C) spinor formalism has been carried 
out as well [3].
Although we also employed the method for demonstrating the 
superradiance in terms of energy and particle number current,
our presentation here is in different context from the existing
ones namely, we work with SO(3,1) Dirac spinor 
and in terms of which the demonstration
appears to be very simple and more straightforward.
In short, our presentation involves
standard formulation of spinor field theory in curved background
spacetime which makes use of Riemann-Cartan formulation of
general relativity in which it is required to
put the Kerr metric given in Boyer-Lindquist 
coordinates in ADM's (3+1) space-plus-time split form in order to
extract soldering form.
This, then, allowed us to show in a very straightforward manner that
the energy current or the particle number current of fermion field
flowing into the black hole through the event horizon exactly
vanishes confirming the absence of superradiance.
Finally, as we stressed earlier, it seems noteworthy that
in the process of illustrating the presence of superradiance in the 
case of the scalar field, almost no reference has been made to the
specifics of the background spacetime geometry, i.e., the Kerr
geometry. As we have seen, we did not need to know even the concrete 
form of the metric of Kerr spacetime. Rather, the result, 
namely the condition for the occurrence of superradiance depends 
on the specifics of the scalar field itself, i.e., its time ($t$)
and azimuthal angle ($\phi$) dependences.
Thus it is interesting to remark on the contrast in the case of fermion field. 
In this time, one needs the specifics of the Kerr geometry to reach
the conclusion on the absence of superradiance. And the result is 
completely independent of the specifics of the fermion field itself.

\vspace{2cm}

\centerline {\bf \large References}

\begin{description}

\item {[1]} R. Penrose, Riv. Nuovo Cimento, {\bf 1}, 252 (1969).
\item {[2]} C. W. Misner, Phys. Rev. Lett. {\bf 28}, 994 (1972) ;
W. H. Press and S. A. Teukolsky, Nature, {\bf 238}, 211 (1972) ;
Ya. B. Zeldovich, J. Exp. Theor. Phys. {\bf 62}, 2076 (1972).
 \item {[3]} W. Unruh, Phys. Rev. Lett. {\bf 31}, 1265 (1973) ;
R. Guven, Phys. Rev. {\bf D16}, 1706 (1977). 
\item {[4]} S. Chandrasekhar and S. Detweiler, Proc. R. Soc. Lond.,
{\bf A352}, 325 (1976) ; S. Chandrasekhar, Proc. R. Soc. Lond.,
{\bf A348}, 39 (1976) ;  Proc. R. Soc. Lond., {\bf A350}, 165 (1976) ;
``An introduction to the theory of the Kerr metric and its perturbations'',
in {\it General Relativity, an Einstein Centenary Survey}, ed. S. W.
Hawking and W. Israel (Cambridge University Press, 1979) ;
{\it The Mathmatical Theory of Black Holes} 
(Oxford, Oxford University Press, 1983). 
\item {[5]} R. M. Wald, {\it General Relativity} (Univ. of Chicago Press,
Chicago, 1984). 
\item {[6]} See for example, N. R. Birrell and P. C. W. Davies,
{\it Quantum Fields in Curved Spacetime} (Cambridge University Press, 1982).  
\item {[7]} R. H. Boyer and R. W. Lindquist, J. Math. Phys. {\bf 8}, 265
(1976). 
\item {[8]} C. W. Misner, K. S. Thorne, and J. A. Wheeler, {\it Gravitation}
(San Francisco : Freeman, 1973).
\item {[9]} S. W. Hawking, Commun. Math. Phys. {\bf 43}, 199 (1975).

\end{description}

\end{document}